*Review*

# Fog Computing Systems: State of the Art, Research Issues and Future Trends, with a Focus on Resilience

**Jose Moura** [1*] **and David Hutchison** [2]

[1] School of Technology and Architecture, ISCTE-Instituto Universitário de Lisboa, 1649-026 Lisboa, Portugal, and Instituto de Telecomunicações, 1649-026 Lisbon, Portugal (e-mail: jose.moura@iscte-iul.pt)

[2] School of Computing and Communications, InfoLab21, Lancaster University, Lancaster LA1 4WA, U.K. (e-mail: d.hutchison@lancaster.ac.uk)

\* Correspondence: jose.moura@iscte-iul.pt

**Abstract:** Many future innovative computing services will use Fog Computing Systems (FCS), integrated with Internet of Things (IoT) resources. These new services, built on the convergence of several distinct technologies, need to fulfil time-sensitive functions, provide variable levels of integration with their environment, and incorporate data storage, computation, communications, sensing, and control. There are, however, significant problems to be solved before such systems can be considered fit for purpose. The high heterogeneity, complexity, and dynamics of these resource-constrained systems bring new challenges to their robust and reliable operation, which implies the need for integral resilience management strategies. This paper surveys the state of the art in the relevant fields, and discusses the research issues and future trends that are emerging. We envisage future applications that have very stringent requirements, notably high-precision latency and synchronization between a large set of flows, where FCSs are key to supporting them. Thus, we hope to provide new insights into the design and management of resilient FCSs that are formed by IoT devices, edge computer servers and wireless sensor networks; these systems can be modelled using Game Theory, and flexibly programmed with the latest software and virtualization platforms.

## 1. Introduction

The digitization and interconnection of almost everything are together making an enormous impact on all aspects of our daily lives. The Internet of Things (IoT) [1] is a very important underpinning technology for Fog Computing Systems (FCSs) [2] that can offer, at the network edge, embedded intelligence and smart actuation/control of peripheral actuators. Prominent examples of an FCS include the intelligent grid, smart buildings, and next generation mobile systems. Because of the increasing importance of FCS in our society, these systems require strong protection against threats that can undermine their normal operation and consequently the quality of our lives [3][4]. Figure 1 visualizes our view of the fast evolution of networked systems towards the emerging FCS. It also shows two design convergence movements. The first of these occurs at the lower part of Figure 1, and is about the convergence of access technologies. The second convergence occurs at the top part of Figure 1, which is related to service access (i.e. remote cloud vs. edge cloud). Our working definition of FCS, particularly the notion of "fog", assumes that the computational resources can be delivered by either remote cloud or edge cloud, depending on each deployment scenario. That is why, in Figure 1, we have the bidirectional arrow between the two top clouds with distinct coverage, and the arrow is labelled with the term "fog". In addition, there are several options for each end device to have access to the Internet. Thus, Figure 1 shows two possible options in terms of access technology, which are Multi-access Edge Computing (MEC) on one side and Cyber-Physical and IoT Systems on the other. Similarly, to the two options for delivery of computational resources, there are also choices for the access technology that will be used for the deployment of a service. We illustrate this second option with the two vertical dotted lines of Figure 1, which are connected by a horizontal bidirectional arrow, also labelled "fog". Another vision of an FCS aligned to ours is in [5], where the low-delay distribution of fog services to end-users uses the cloud-to-things continuum infrastructure. This vision is equivalent to the top part of Figure 1, associated with the resources (fog vs. edge) used to compute the tasks of each service. The FCS vision that we depict complements the vision in [5]: here (in Figure 1) we are adding the new aspect concerning the heterogeneity of network access.

The previously referred FCS threats can be grouped in two classes, viz. unplanned and planned. The unplanned system threats are typically due to natural disasters (e.g. earthquakes) or non-intentional faults or human errors. Planned system threats are typically associated with cyber-attacks. Whatever the origin of the threat, it is vitally important to deploy appropriate resilience strategies and mechanisms to mitigate threats. A list of relevant system weaknesses against threats is available in [5]. This list includes low-level vulnerabilities in physical networking equipment using field programmable gate array (FPGA) devices, which may allow, as an example, an attacker to install and boot a malicious software image in a huge diversity of networking devices, such as routers, switches and firewalls [6].

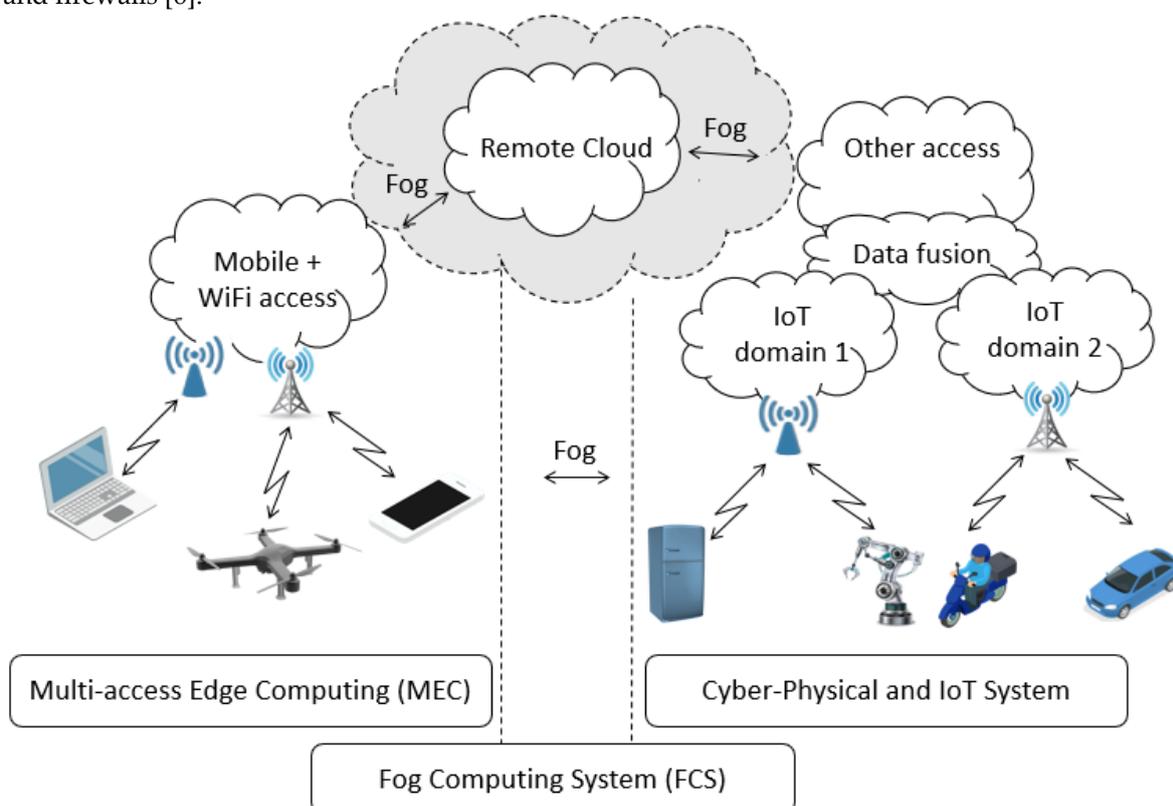

**Figure 1.** Evolution of Networked Systems to Fog Computing Systems

FCS is about a physical facility with embedded sensors and actuators that can be remotely monitored and controlled by computerized systems [7]. The monitoring and control are made by logical control loops over physical communication channels. These channels are established between the sensors/actuators and the computers that manage them. The channels transfer both data representing the facility status and control messages to change the operation mode of that facility. FCS threats can adversely impact both system monitoring and control tasks.

The Oxford English dictionary defines resilience as "the capacity to recover quickly from difficulties". This suggests that a resilient system in the face of some either known or unknown threat should contain features to mitigate or even prevent that threat; typically, these features will be to detect, absorb, recover and adapt [8]. After successfully detecting a threat, the system should have the 'absorb' feature to diminish the negative impact induced by that threat. Following the threat occurrence, the system should recover its operation as quickly as possible to an acceptable level. Then, after the system threat has finished, the system should adapt its management policies and diminish even more than before the negative impact on the system in any future repetition of that threat or even prevent it. During the current paper resilience and robustness are considered as synonyms. Nevertheless, [9] differentiates these terms. They argue that, after a system threat, a resilient system shows a temporarily degradation on its performance. Alternatively, a robust system does not have any degradation. In addition, [9] discusses attributes closely related to resilience such as reliability, agility, flexibility, among others, which are not covered here. We also consider that resilience is related to both physical system infrastructure and data (logical) analysis [7].

The high levels of heterogeneity, complexity, and dynamics of resource constrained FCSs bring new challenges to their reliable operation, which imply the need for novel management strategies, using distinct technologies. These technologies are Game Theory (GT), Software-Defined Networking (SDN), Network Function Virtualization (NFV) and Machine Learning (ML). As an example, GT is a fundamental tool for modeling the threats and their interactions with these systems, enabling the design of automated protection mechanisms [10]. GT has been used to study mechanisms in the area of Advanced Persistent Threats (APTs) [11] as well as for enabling both safety and security in cyber-physical and IoT systems [12]. In the context of an APT, a well-provided resource attacker establishes a long-term, illegal, and obfuscated infiltration in a network domain to steal confidential data. GT can model and analyze the interactions between threating entities and system defenders to protect the data privacy and enable the resilient operation of FCSs.

The emerging FCS features [8] of detect, absorb, recover, and adapt are like the diverse stages of a theoretical game model that runs in a sequential way, mediating the interactions among opponent players. Examples of opponent players are cyber-attackers versus automated defense mechanisms, or nature versus self-healing mechanisms. We argue that an effective way to fulfil the challenging requirements imposed by the successful management of resilient FCSs is to orchestrate diverse technologies, such as GT, SDN, NFV, and ML.

Previous work in the security area has investigated privacy on IoT systems [1], and safety [12] or resilience [13] for Cyber-Physical Systems (CPSs) with embedded IoT devices. Our contribution in this paper is to refresh and advance the literature by investigating effective management solutions to the novel requirements of a new generation of Cyber-Physical Systems (CPSs) augmented with other technologies, such as IoT. All these technologies are converging into a common platform, often labelled as Edge (Fog) Computing. This new version of CPSs we call FCSs. In the current work, we discuss the available literature, highlight the advantages that the surveyed research areas could bring to FCSs for enhancing their resilience, and outline some future research areas. As a primary case study, we investigate how an FCS system can detect, absorb and recover from, and adapt to, threats. These goals should be achieved using available system resources for limiting the system deployment and maintenance costs. Despite these considerable constraints, FCS should fulfil time-sensitive functions with variable levels of integration with the environment, integrating distributed data storage, parallel computation, proficient communications, ubiquitous sensing, and efficient control of system resources used by machines, agents (algorithms) or end-user devices. Our work is mainly concerned with FCSs. However, the entire discussion throughout the paper is by no means restricted to FCSs. In fact, the main findings of our work can be also applied to more general scenarios, such as wireless/IoT networks.

This paper has the following structure. Section 2 discusses related work. Section 3 describes some background to the remaining part of paper, namely pertinent scenarios in FCSs and relevant aspects to be aware of when the goal is to enhance the resilience of FCSs. Section 4 presents a four-layered design for resilient Fog Computing and IoT systems. Section 5 outlines open research directions. Section 6 concludes the paper. The paper's logical organization is visualized in Figure 2.

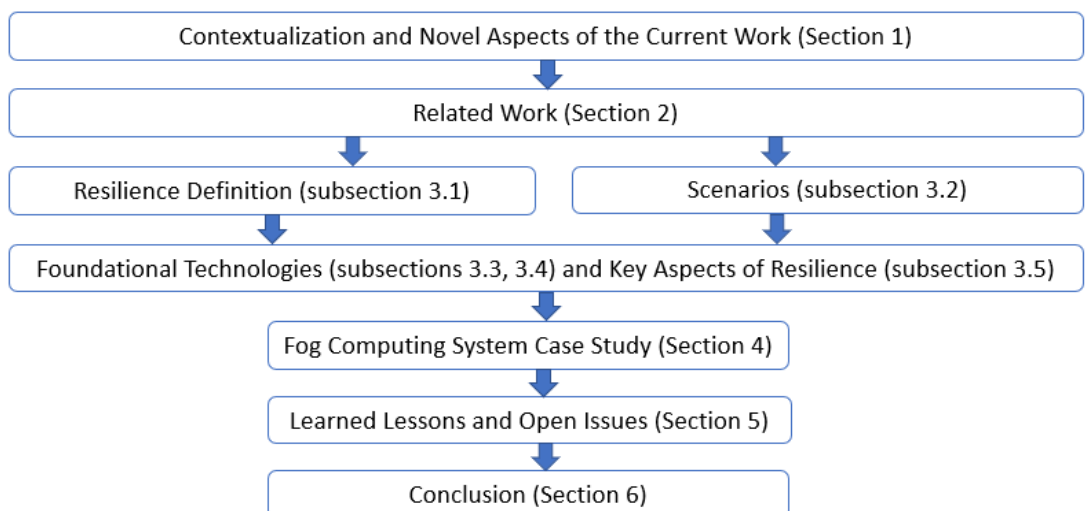

**Figure 2.** Logical Roadmap Behind the Paper

## 2. Related Work

This Section compares known related work at the time of writing. We discuss relevant associated technologies, including Game Theory (GT), SDN, NFV, and Machine Learning (ML). We also envisage GT as complementary technology to the other technologies discussed in our paper. All these technologies could have an important role in efficiently managing fog computing systems.

In our view, the technologies mentioned above will be of paramount importance to support new applications that will appear in the next decade. These emerging applications will demand much beyond the best-effort service offered by the current Internet. These new services need the Internet to offer strict packet latency, to provide exact coordination among many packet flows across multiple communication channels, and to assure connectivity service immune to the negative impairments of any Internet threat. This paper is focused on the last of these capabilities, viz. resilience, in the context of Fog Computing. We have surveyed the relevant literature as discussed below. However, first a clarification about the network/computing architecture scope of fog computing systems: our architectural view of a FCS is like a bidirectional convergent movement of technologies and computational resources (see Fig. 1) not only from the cloud to the edge devices, i.e. the decentralized cloud discussed in [14], but also from the edge devices to the cloud, which was debated in [15]. That is, the fog edge is a moving border, depending on the perspective initially analyzed (i.e. cloud or mobile ad hoc access).

From the literature analysis, we can conclude that a vast amount of work, e.g. [16][17][18] has been done on studying and classifying published contributions in the security area, but they lack a more holistic and complete revision than the security perspective by reviewing some other important areas, such as software-defined solutions for resilient FC Systems (FCSs) operating in diverse emerging use cases. Considering that resilience incorporates security as a sub-capability, we justify the classification of some of the surveys as constrained literature overviews, because (for example) some [16] present techniques for detection of DDoS attacks, while others [17] study machine learning solutions for intrusion detection. Also, [18] is focused on security control and attack detection but only for industrial FCSs. Further relevant and associated work [19] is focused on theoretical management models for networked systems, considering the need of the diverse players of the network to cooperate among themselves to enforce the optimum usage of the available network resources. The authors of [13] review dependability and security for detection, diagnosis, and mitigation. Nevertheless, they do not cover enabling technologies, nor applications, in their discussion. Clearly, further work is necessary for investigating proposals which can enforce not only a particular security aspect but a more generic resilience framework to immunize network operation against the negative effects of serious threats.

The authors of [20] argue that the management of networked systems should also have self-adaptability characteristics in the presence of serious system threats. In this way, the system structure as well as its operation can become more resilient against serious and persistent menaces against their normal operation. Aligned with this set of system self-capabilities, there are a set of work proposing ML techniques [17][21][22]. In a more detail way, [22] investigates how ML algorithms can be used in applications involving NFV and SDN. NFV virtualizes network functions and decouples these from the hardware. The main goal of NFV-based solutions is to automate network configuration as well as to provide system services in an elastic and adaptive ways. On the other hand, SDN can be very useful in Edge Computing scenarios to program, with some abstraction of the physical devices, the way the networking-based system is expected to operate.

SDN-based solutions can divert computational-intensive tasks from resource-constrained mobile devices to more powerful servers located at the network edge. Hence, the battery autonomy of mobile devices is increased, and the results of the computational-intensive tasks are more quickly obtained. In addition, there are several important contributions also aligned with our current proposal but applied to a more specific use case, namely, network resource allocation for ultra-dense networking [23], mobile network planning with small cells [24], optimization of hybrid SDN networks [25], and management of faults in SDN-based [26] or vehicular networks with autonomous cars [27].

In addition, we have found valid contributions in sensing management for smart city monitoring [28], mobility of things or devices using several standards of low-power wide area networks [29], studying novel business models for resource management in 5G wireless networks [30], and analyzing datasets to predict and prevent security incidents [31].

From all the surveys we are aware of at the time of writing, we clearly identify [32], which broadly reviews the literature in terms of contributions to resilience applied to FCSs. The authors argue according to their experience that the most challenging issue for designing a resilient FCS is to deploy a real-time, closed-loop, networked control system completely immune against serious threats. These threats are caused by natural noise, which induces in the system packet loss and bit errors, as well as some internal or external cyber-attacks. They also discuss two medical case studies. The first is about the resilient integration of virtual reality and a robot device for restoring the corporal coordination and flexibility of persons with disabilities. The second study is about the design of a robust implantable medical device for physical-to-cyber health sensing and cyber-to-physical organ control. This survey published in 2016, it is naturally outdated, missing last relevant and related literature work. Our paper tries to overcome these shortcomings, and looks anew at the literature on reliable and flexible software-defined management solutions for resilient FCSs.

The current work adopts a set of management features [8] (i.e. detect, absorb, recover, and adapt) to support the resilience of FCSs. Nevertheless, there are other alternatives, such as in [4] where the authors propose two types of management features; the first type is formed by short-term or reactive management activities, viz. defend, detect, remediate and recover. The second type is formed by long-term or proactive management activities, viz. diagnose and refine.

The next Section offers a background discussion on these technologies, as well as on FCS scenarios and resilience requirements in systems formed by either physical or virtualized resources.

## 3. Background on Software-Defined Resilient Fog Computing Systems

The paper's content is based on our outline of a software-defined resilient Fog Computing System shown in Figure 3. It shows the two major goals of the paper. The first goal is to investigate how GT can model an FCS with IoT applied to edge computing use cases. The second goal is to study how SDN-based solutions can empower FCSs with embedded IoT devices communicating via wireless sensor networks with extra capabilities, such as programming of management policies that enforce the system resilience against diverse threats, including system node failures. For that, our current view is about SDN-based solutions that orchestrate other technologies such as NFV or ML. In this way, there is a viable solution to support dynamic cross-layer resilience decision making for FCSs using heterogeneous IoT devices. In the subsections below, the literature within the major areas evidenced in Figure 3 are discussed, as follows: i) resilience definition and background; ii) scenarios of Fog Computing Systems (FCSs); iii) analysis of FCSs; iv) novel programming models for FCSs.

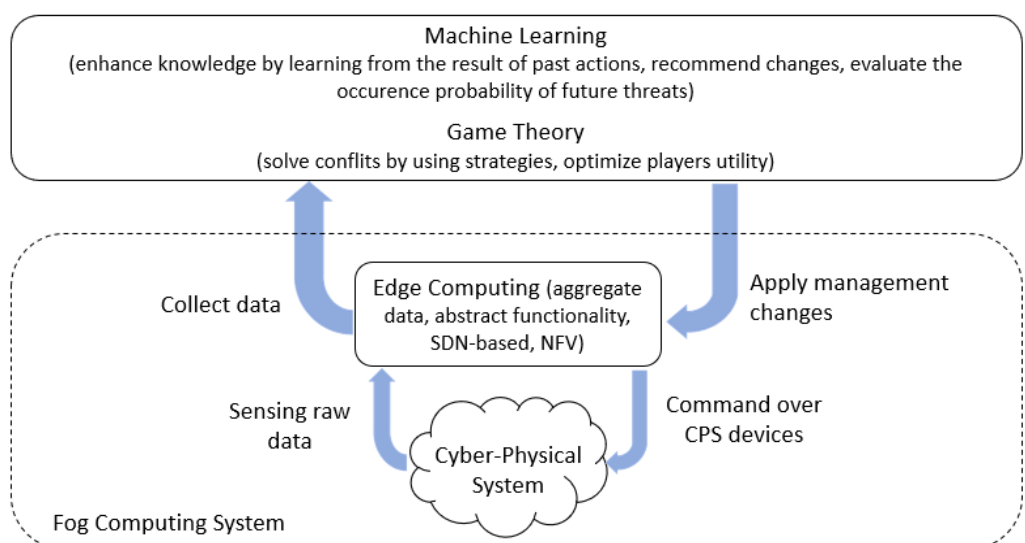

**Figure 3.** Outline of a Software-Defined Resilient Fog Computing System

*3.1. Resilience Definition and Novel Management Challenges*

The NIST[1] definition of a Cyber-Physical System (CPS) is a commonly accepted one. According to that definition, CPSs are engineering-based systems offering a functionality strongly dependent on the interaction among computational and physical processes. This integration enables the deployment of emerging systems that our society can use in different ways. The literature also offers several definitions of resilience. Our paper adopts the definition of resilience as the system's capability to maintain an acceptable level of service to its users despite the eventual occurrence of various faults and challenges to the system's normal operation, where some issues could be completely new to the system architect [4]. So, a resilient fog system should offer a satisfactory level of service despite the various challenges to which it is exposed, whether these be natural disasters, weather events, component failures (in hardware or software), misconfigurations / human errors, or malicious offensives such as cyber-attacks.

FCSs show completely novel capabilities, including pervasiveness and intelligence. In parallel with the new benefits offered by FCSs, these systems also become more attractive to cyber-attackers because if the attacks are successful then important sectors of society could suffer significant losses. Considering security as one of the relevant aspects of resilience, we have found several publications that discuss the use of public-key solutions in secure fog computing scenarios, e.g. involving IoT devices [33][34]. We think that public-key solutions based on hierarchical centralized design (e.g. Public Key Infrastructure – PKI) can have scalability issues, robustness problems due to their single point of failure, or be especially attractive to cyber-attackers due to their centralized operation. In addition, given the limited computational resources in edge/fog/IoT scenarios that we recognize in this paper, it is generally difficult (and thus sometimes neglected) to apply public-key crypto in FCSs, sensor networks or other systems. But these security services need to be used, without doubt. At this point, a pertinent question arises: are there good solutions to handle the costs of cryptographic certificates in FCSs? The research community surely has an important role in successfully answering that question. For this, GT (discussed in subsection 3.3), the novel programming approaches (debated in subsection 3.4), or our hierarchical design of a resilient FCS (discussed in Section 4) may eventually be useful to answer this question. In addition, we have found some alternatives to PKI models with a more decentralized design [35][36]. Clearly, further work is required in all these security aspects to successfully deploy them in FCSs.

---

[1] NIST stands for National Institute of Standards and Technology, available in https://www.nist.gov/ (verified in 21/02/2019)

The research community has dedicated recent notable efforts to address the novel threats against FCSs [18] by enhancing the management of resilience [37][38]. In addition, the relevant interplay between resilience and self-organization for the design of critical networked systems has been also investigated [20]. A potential outcome from all these efforts is to obtain a robust FCS, which should have a system design oriented for the aspects of stability, security, and 'systematicness' [32]. By 'systematicness' they mean a system that has a seamless integration of sensors and actuators. Please consult [32] for further details on these design aspects. According to an ITU recent deliverable [39], the resilience is a very relevant network capability to maintain the high quality, availability, and reliability of the upcoming networking services and their related user applications.

The literature have a few contributions related with FCSs that immunize these systems against either already identified or potential threats in distinct usage scenarios. The next subsection presents a literature revision of FCS using a taxonomy based on some relevant use cases.

*3.2. Scenarios of Fog Computing Systems*

We start this subsection trying to clarify the Fog Computing (FC) paradigm. The authors of [40][41][42] consider that FC is not a substitute for Cloud Computing but a powerful complement. In this way, FC enables elastic processing at the network edge and low-latency user access to the data output from that processing, or even data stored at the network periphery. Nevertheless, it is still possible, if necessary, to connect remote cloud applications to the end-users. In addition, FC serves as a key enabler [43] for many future technologies like 5G [44][45], Internet of Things (IoT) [46][47][48], Blockchain [49], or even applications requiring computation offloading [50]. The FC paradigm besides providing local processing/storage and low latency, it also supports mobility and location awareness for specific applications requiring those requirements for their normal operation.

Then, the current subsection discusses below some relevant use cases in Fog Computing Systems (FCSs) enhanced with the ever-developing model of the Internet of Things (IoT). It provides a concise and precise description of these scenarios, their basic requirements, and the novel challenges that these use cases present to the research and standardization communities. The scenarios discussed below involve power grids [51], smart buildings [52], next-generation mobile communication systems [53], healthcare systems [54], and Industry 4.0 (Industrial IoT) [55].

Intelligent Power Grid Systems

Power systems are changing from centralized large facilities to fully distributed micro-size facilities, such as domestic electrical appliances composed by photovoltaic panels and DC/AC power inverters. At the same time, the full operation of the latest generation of microgrids depends massively on computerized systems. Consequently, that situation makes these systems increasingly exposed to cyber-attacks [51][56], which could create huge problems in our everyday lives.

Smart grids present great challenges to their efficient management due to the unpredictability of demand load and the reliability of data communications. The unpredictability of demand load is caused by many factors, not least to domestic electrical energy variation and most recently to the charging of electrical vehicles [57]. In addition, the reliability of data communications in smart grids could be adversely affected by perturbations that occur on some links, mainly on wireless links [58]. These perturbations can be induced by either cyber-attacks [59][60] or impairments on the communications medium [61][62][63]. The authors of [61] study the outage probability of a wireless link, considering the multipath fading, shadowing, and random path loss given the location distribution of smart meters. In [62], they prioritize the data transfer within a smart grid using a position-based quality-of-service (QoS)-aware routing protocol. Further, they propose a load-balancing mechanism to mitigate network congestion induced by many critical event messages. As an example, these event messages can be related to the high number of damages on the electrical physical infrastructure a big storm can cause.

Paper [64] proposes an incentive-based demand response algorithm for smart grids, which uses a deep neural network to overcome system uncertainties by finding out the initial unknown both prices and energy demands. In addition, the same algorithm also uses reinforcement learning to obtain the optimal incentive rates, considering the profits of both service provider and customers.

In a smart grid, several devices use wireless communications to transfer data. Most of these use industrial, scientific, and medical (ISM) radio band for channel communication. Since the ISM band is license-free, attackers can easily have access to that frequency band, trying to initiate a cyber-attack. A potential attack is the jamming attack that can disconnect important system devices such as smart meters, collectors of meter data, remotely controlled distribution automation devices, and GPS antennas of phasor measurement units. Therefore, more work is needed to create resilient wireless communications among the diverse components of a smart grid.

The authors of [65] are concerned with the resilience of network control systems under communication delay attacks. These attacks force two wrong operational situations on the controlled system (e.g. power plant) due to either missing data or delayed data from the power plant status. In addition, this attack type can be made over encrypted messages by jamming the wireless communications used by the system control loop. So, the authors of [65] propose a solution to countermeasure a time delay attack. To implement that protection, their solution has a system state estimator. Then, the state predicted can be compared with the reported state of the power plant. In the case the two states show significant changes, then the controller uses the predicted state until the reported state is similar to the predicted one, when the controller uses again the reported state.

Further aspects that need more work in smart grids are anomaly detection systems and intrusion detection systems, particularly from insider attackers [51]. In addition, more investigation on coordinated cyber-attacks needs to be carried out. During a cyber-attack bogus data that is enforced by the attackers can affect the normal operation of a power plant. So, it is necessary to detect and remove as quickly as possible from the power grid supervision system all the bogus power plant status inserted by false data injection attacks [66][67]. The authors of [68] provide a detailed discussion of improvement strategies for the resilience of power systems. These strategies are classified based on two distinct perspectives. The first perspective analyzes the resilience of power systems considering the time-dimension. The second perspective enforces the robust operation of power systems, choosing adequate control actions. Additional discussion on these topics is in [68].

Smart Building Systems

Considering the substantial price reduction in sensor nodes, these now can be used in novel applications, as is the case, for example, in modern smart buildings. The authors of [52][69] argue that IoT can be a catalyzer of a full integration of building intelligent mechanisms with the system grids that connect each building. Hence, the people living within those buildings could benefit from greater comfort without paying more to the diverse utilities, e.g. electricity, gas, water, or even healthcare [69]. To achieve this, and particularly for the electrical scenario, all the intelligent control mechanisms existing within a building should operate in a completely coordinated way with the smart power grid supplying that building. To make that coordination possible, open networking protocols should be used [52]. By open, we mean standard solutions that enable universal exchange of data over heterogeneous technological systems to fulfil a set of common goals.

The authors of [70] propose a population based algorithm for deciding about the places where sensors should be located to monitor the various pipes within a large building. Their results show that the proposed algorithm demonstrates good performance in relation to other algorithms inspired in nature. The results obtained also suggest that the system lifetime can be improved.

In [71] the authors propose a smart solution for buildings that learns and predicts optimum individual user-preferences towards the efficient energy control of personalized light. They argue that their proposal can achieve energy savings up to 72% when compared to the conventional lighting systems. Aligned with previous work, [72] proposes a flexible approach supported by deep learning that offers automatic adjustments to system / environment variations. The same approach also has an incentive mechanism based on gamification for improving the interaction between the building inhabitants and the building system that supervises and controls the infrastructure.

In [73] the authors propose a platform-based methodology for smart building design. The last platform reuses hardware and software on shared infrastructures, enables the fast prototyping of applications, and allows exploration of the design space to optimize the design performance. The paper illustrates the usage of the proposed platform via a case study on the design of on-demand heating, ventilation, and air conditioning systems.

More work is necessary in the area of smart building systems. In fact, the living comfort characteristics (temperature, humidity, air quality), energy (electrical or gas) efficiency, and building safety require the existence within each building of a software-defined management service responsible to satisfy the new challenges imposed by modern buildings with a myriad of embedded sensors and actuators. This management service should be responsible for local control loops, send automatic messages to external public entities (e.g. in case of gas leakage or fire alarm), and recording the occurred building events for future analysis. After analyzing the data extracted from the building events, the management service, in a proactive way, could recommend some maintenance tasks in the building to improve the living comfort, reduce the energy consumption, increase the safety, or reduce the false-negative statistics of reporting failures.

Next Generation Mobile Systems

Considering the rapid evolution of mobile cellular technologies, including smart personal wireless devices, a set of new mobile applications is appearing [53]. These novel applications are mainly focused on fulfilling the requirements of users. For ensuring more positive usage experiences to end-users, service providers are moving their focus from Quality of Service (QoS) to Quality of Experience (QoE) in the way they aim to efficiently manage the available resources from their network infrastructures. Obviously, QoS is related with technical metrics such as packet loss, loss rate, delay, and jitter, whereas QoE try to balance in the more positive way what the user expects from the network-based application and what the same user effectively gets from that application. In addition, one can see QoE [53] requirements as an evolution from the those of QoS [74][75]. For next generation mobile systems (e.g. 5G), the authors of [53] comprehensively discuss the literature in terms of enhancing the user experience by means of supporting advances in the methods that assess the video quality and reflecting on how the QoE reported from users should be conveniently managed in upcoming usage scenarios. For further enhancements in QoE, the management of both network resources and offered services needs to be evolved by adopting solutions based on self-organization optimization. In addition, SDN and cloud technologies can be useful to allocate the required services to the best possible available system resources, enabling a more dense network management with smaller cells than before, and a more holistic management considering the cross-layer aspect of SDN operation [76]. The authors of [77] propose a potential game for sharing spectrum in 5G networks in a decentralized way and based on user QoE. The work in [78] discusses the major challenges and future developments on FCSs in vehicular networks, healthcare systems, and mobile education. The contribution in [79] considers the scenario of a vehicular ad hoc network formed by vehicles on the road with some common interests which can form a platoon-based driving pattern. They comprehensively discuss the novel management challenges induced by the platoon-based driving in the efficient operation of the vehicular network.

Further work is needed to manage the continuous convergence between the network operators and the cloud providers forming a common meet-in-the-middle place currently designated as edge (fog) computing. This convergence scenario is a win-win situation for all the players involved, including the end-users, as we now explain. The network operators need more services, processing, and storage resources from the cloud providers, including their experience in satisfying high volumes of data processing with a minimum set of computing/networking resources by orchestrating all these (physical/virtualized) resources in an elastic way. In the opposite direction, the cloud providers are interested in supplying the end-users with customized applications and with the highest possible quality. To satisfy these requirements, the proactive data caching at the network edge based on historical data popularity can be very useful to diminish the data access latency/jitter. End-users want ubiquitous and reliable access to all the services they need in each of their daily tasks. If the end-users are served adequately then they will be satisfied, rewarding both network operators and service providers. Here, novel business models can give the right incentives towards the goal of being offered applications with a quality much higher than the current best-effort model. The network provider should receive the adequate incentive to supply the user with the right amount of network resources. On the other hand, the network should have an appropriate accounting mechanism to verify the exact amount of network resources that have been used by each user application. This is to ensure some fairness among the diverse users' payments. GT can enhance such business models by solving eventual conflicts among the self-expectations of the diverse players.

Healthcare Systems

A large and rapidly growing percentage of people in most countries is elderly. There is huge pressure to devote enough medical and human resources to ensure a good quality of life. However, the commitment of enough resources is proving impossible by reason of both human and financial constraints. A popular approach to alleviate these pressures is to explore the adoption of IoT in medical service systems, enabling innovative solutions in healthcare [54]. There are important potential advantages of deploying IoT-based healthcare systems, namely: i) extract useful information for raw-data; ii) automation in terms of either improve patient health or promote preventive care; iii) enhance patient satisfaction and engagement with their treatment procedures; and iv) enhance the management of population health in a large scale with a suitable amount of resources. However, obstacles for adopting IoT in healthcare systems include issues with security and performance. As an example, for deploying IoT-based healthcare systems with excellent performance, there is a strong need to support real-time requirements. The authors of [80] propose a fog computing implementation to decrease latency substantially. This occurs because the data processing is made as close to the end-consumers as possible by leveraging virtualized containers on the network edge such as mobile base stations, gateways, network switches and routers. Additional deployment challenges of IoT-based healthcare systems are discussed in [81]. In addition, based on a thoroughly revision of the literature, [82] discusses the major advances healthcare systems embedding IoT-based smart devices. They also address the intelligent trend and future research directions in the field of IoT-based healthcare solutions.

The authors of [83] proposed a smart health system which includes a unified data collection layer for the integration of public medical resources and personal health devices. In addition, the same system has a cloud-enabled and data-driven platform for multisource heterogeneous healthcare data storage and analysis. Then, the system offers access interfaces for system developers and users.

A comprehensive analysis of authentication protocols which address the trade-off between securing implantable medical devices in terms of access rights and the safety of the patient in case of emergency is available in [84]. Moreover, they contrast the authentication protocols with respect to the cryptographic and security mechanisms implemented on the implant.

Healthcare applications can benefit from the deployment of the fog computing paradigm [85]. This could be aided by appropriate design innovations in the way the networked systems interoperate [86]. In addition, healthcare applications can offer low latency, distributed processing, context awareness, better scalability, fault tolerance, better security and privacy [80]. Cloud-based IoT systems [87], or fog-based [85] to reduce the access delay, seem to be promising solutions in healthcare due to the huge available capacity in data storage and data processing; the offloading of computationally intensive data analysis tasks from body sensor devices to fog servers, which could be containers, is feasible. In this way, the autonomy of battery-operated body sensors is increased, lessening the burden on patients as they will not so often need to recharge their body sensors. The use of remote-location healthcare has also become viable with the recently proposed Internet of Medical Things (IoMT) [88]. The IoMT approach can enable the diagnosis and treatment of patients to give a higher level of satisfaction, especially during pandemic periods or even in cases of patients with mobility impairments. A taxonomy of FCSs for healthcare is available in [89]. They identify as open research challenges the issues related with security and privacy, autonomic decisions for loop control, and event prediction.

Industry 4.0

The term "Industry 4.0", the fourth revolution in the field of manufacturing, was coined in November 2011 under a governmental initiative to enhance the German competitiveness in the manufacturing industry [90]. Industry 4.0 is also more recently designated as "Industrial Internet of Things (IIoT)" or "Smart Factories". It is a new sort of industrial revolution that not only guarantees communication and interconnection among distinct industrial systems, but also analyzes the information obtained from it, and use that information to create a more holistic and better-connected ecosystem for the industries [55]. In addition, the automation offered by the deployment of IIoT makes the industrial plants more autonomous and efficient, but it is still possible to have remotely operated industrial processes. However, the capabilities of reliability and stability for critical communication, with short and predictable latency, are required to offer remote services to these Industrial factories. In addition, to diminish the deployment cost of the communications infrastructure inside each industrial plant, the communication should be supported by wireless networks. In this perspective, 5G is a superb candidate to support the communication in Industrial networks, because it supports three essential communication types, i.e. deterministic low-latency, massive machine-to-machine communication, and enhanced mobile broadband [91].

It is expected that ultra-reliable wireless communications, supporting massive communications among the industrial machines, devices, and robotized actuators, will enable the development and transformation in smart factories, while contributing to the growth of social and economic aspects. However, the engagement of many sensors and other industrial equipment will increase the amount of data to be captured, stored, and processed. This pervasive and exponentially increasing wireless data traffic is usually characterized as "big data" [92]. Here, an SDN-based architecture combined with a "big data" engine can offer a data-driven intelligent networking infrastructure to enable the processing of massive amounts of data to obtain useful knowledge about each Industrial process.

The data-intensive transformation expected to occur in future industries will generate not only large amounts of data to be processed but also most of those data are not organized in a clear structure. All these facts bring a huge difficulty to humans who process this data in order to obtain useful information to supervise and manage the industrial processes. A possible solution for the last problem is to create an autonomous data-driven decision making system, using SDN, NFV, and ML [93]. In addition, ML can be also used to enhance the wireless broadband access in order to the strict requirements imposed by industrial communications can be supported in a satisfactory way [94][95].

The role of human-computer interaction in future industries will also incorporate virtual/augmented reality enhanced with haptic technologies [55]. These haptic technologies provide the remote human operator with a true immersive experience on the industry plant's physical context, with the final goal of obtaining a more efficient remote control of the plant. The haptic information is based on sensors typically physical object harness, weight, inertia, and positioning awareness, as well as surface contact geometry, smoothness, slippage and temperature [55]. In this context, the network needs to support a new type of haptic communication that transmits to the remote operator correct and complete feedback about the industry plant status. In this way, the network should offer a medium to transfer in real-time the sense of touch (haptic) and actuation (kinesthetic). Other relevant requirements of these forms of communications include low latency, high throughput, reliability, and intelligent precise coordination (synchronization) among the diverse flows [96].

Summary of Open Issues in Analyzed Scenarios of Fog Computing Systems

Table I summarizes some prominent functional aspects from the analyzed FCS scenarios that require further work from the associated research community.

**Table I. Summary of Open Issues in Analyzed Scenarios of Fog Computing Systems**

| FCS Scenario | Open Issue |
|---|---|
| Intelligent Power Grid Systems | Anomaly detection from insider attackers; protection against coordinated cyber-attacks; immunization against data injection attacks in the control loop |

| Smart Building Systems | Deployment of a software-defined management service to enhance the living comfort and safety as well as a proactive recommendation service for building maintenance tasks |
|---|---|
| Next Generation Mobile Systems | Proactive data caching at the network edge based on several data dimensions: popularity; spatial; temporal; energy to data storage, processing, or transfer. |
| Healthcare Systems | Offloading of computationally intensive data analysis tasks from battery-operated embedded devices to fog servers aiming to enhance the battery autonomy of those devices |
| Industry 4.0 (IIoT or Smart Factories) | Ultrareliable wireless communications, supporting massive data communications; autonomous data-driven decision making; haptic communications |

*3.3. Modeling of Fog Computing Systems*

The modeling of a Fog Computing System is made using a theoretical model. A very popular tool to model and perform system analysis is Game theory (GT). It is a fundamental mechanism to study the various challenges and faults that could affect the system normal operation. GT can also enable the design of automated mechanisms to protect the major functionalities of the system [10]. Nevertheless, theoretical game models may need a significant amount of time for discovering stable and optimum system configurations. This extra delay introduced by GT may not meet the low latency needs of fog systems. We think one way to cope with that problem is to deploy the theoretical model running in the backend of the fog system, where the theoretical model acts as a northbound SDN application (SDN is discussed in subsection 3.4). In this way, the theoretical model has an active role in the system operation only in those instants of time where the model potentially converges to a viable system configuration. When the theoretical model is running in the backend it will be the SDN controller's role to manage online the CPS's context.

Literature Review for Theoretical Models

We have reviewed the literature for theoretical models in Fog Computing Systems distributed in the following aspects: system resource allocation [19][23], system resource offloading [97][98], system energy efficiency [72][99], and system security [100][101].

A brief background in GT (non-cooperative, cooperative, and evolutionary games) applied to edge computing is available in [19]. It includes a comprehensive review of game theoretical contributions to wireless communication networks. They also discuss diverse issues that can be addressed by theoretical game models to optimize the network performance in some emerging multi-access edge computing scenarios. The authors of [23] offered an alternative literature revision in coalitional games among other alternative techniques, such as large-scale convex optimization, mean field game, stochastic geometry, and stochastic optimization. All these techniques can enable an optimized system resource allocation. In particular, the mean field games can be applied to analyze scenarios with many resources, devices and user types. The authors of [63] proposes a non-cooperative game theoretic model for the management of a smart grid's demand considering the packet error rate in the game formulation. In [102] is the authors describe a model for determining optimal resource allocation by combining GT with a multi-attribute utility model. It allows optimal allocation of the defender's budget across potential identified system targets and, considering different types of countermeasures. The authors of [103] propose a hierarchical model between mobile operators and users. They offer a management solution for effective bandwidth slicing in software-defined 5G systems.

Other relevant functional aspects in FCSs are data offloading [97] or computation offloading [98][104][105]. In [97] the authors outline a SDN-based controller enhanced with a game model based on a single leader and multiple followers for a 5G-based vehicular network. This solution aims to deal with high speed and traffic congestion within that network. It enables the vehicles to perform intelligent decisions for data offloading by using the network services of priority manager and load balancer, which route the traffic load in an efficient way even within a large network. Another contribution [98] considers a scenario of a heterogeneous cellular network. The authors propose a solution based on a two-stage auction to perform task transfer from macro cell users to small cells, which are relay network nodes for task execution. This alleviates the heavy burden of macro base stations by offloading computation from macro cell user equipment to small cell base stations or remote cloud. The work in [104] offloads user processing tasks from the user devices to the network edge. Then, at the network edge the offloaded tasks are scheduled between the remote cloud and edge servers. This task scheduling is performed by an evolutionary algorithm, which aims to protect the quality needs of applications and their computation-intensive resources. In these cases, the available computational resources in MEC servers may not be enough to fulfil the computation delay, and consequently the tasks need to be processed in the cloud servers. In [105] the authors propose a non-cooperative and distributed game among Industrial IoT devices with the assistance of Blockchain. Using Blockchain, the IoT devices could securely trade distributed resources with other untrusted peers. Their solution transfers heavy resource demanding tasks such as data processing as well as mining from IoT devices to the edge/cloud servers.

There are some literature contributions focused in either reducing the energy consumption [72] [99] or defeating jamming attacks [106]. The authors of [72] propose a social game aiming to incentivize building occupants to modify their behavior so that the overall energy consumption in their room is reduced. Aligned with this, [99] proposes a social game to guarantee energy efficiency for buildings. Interestingly, [106] uses energy harvesting as a countermeasure against a potential jamming attack. The harvested energy is extracted from the energy used by the attacker to jam the channel, and the former is consumed to increase the transmission power of benign traffic.

A significant amount of work was made to enhance the security of FCSs. The authors of [100] classifies the literature into two classes, viz. security and privacy. Then they discuss the work in terms of the GT model that each contribution has proposed to defeat the various cyber-security problems. The authors of [101] propose a hierarchical model that adjusts the strategies for enabling the selection of wireless channels in such a way that jamming attacks are avoided. The work in [107] investigates how a game-of-game concept formed by two intertwined games can study the tradeoff between robustness, security, and resilience of a cyber-physical control system. The first game is a zero-sum differential game for robust control design at the physical layer. The second game is a stochastic (explained below) zero-sum game between an administrator and an attacker for the design of an effective defense mechanism. By using this game-of-game framework, the authors aim to defeat the potential threats originated by attackers exploring the vulnerabilities not only from the physical part (i.e. the physical plant and control layer) of the system but also from the cyber elements (i.e. communication and network layers) that enable advanced system monitoring and intelligent control. Another contribution [108] proposes stochastic games for protecting microgrids against cyber-attacks. In a stochastic model a dynamic game is played with probabilistic transitions and in a sequence of system steps. Such games start in a given state, and then the players select actions according to their own status at the time of each system phase. The players receive an immediate payoff according to their status and the actions selected in a probabilistic way. This procedure is repeated through a sequence of system steps with each player aiming to obtain the maximum total utility from the interaction with the system and with others. The authors of [109] apply a dynamic Bayesian game of incomplete information to implement cyber deception by means of Honeypot devices in IoT usage scenarios. Honeypots are virtual or physical network devices that emulate exact network nodes to attract to them potential malign traffic. In this way, two positive outcomes for the network security are guaranteed. The first one is to starve the resources and time of the attackers by deviating the bad traffic from the initial targets of the attacks, i.e., the real network nodes, which become protected from the negative impact of those attacks. The second advantage is to analyze the behavior of each attack, including a completely new one, learning from the analyzed trend of each attack how to develop future effective countermeasures to mitigate that attack.

Comparison among Modeling Techniques for Fog Computing Systems

Table II compares FCS modeling techniques in terms of their strengths and weaknesses. The modeling techniques under comparison, in our opinion, are relevant for discovering suitable management decisions to counteract the negative effects of various threats to the normal operation of an FCS. In addition, [110] discusses further theoretical games in the network/cyber-security domain, such as, trust assignment, resource allocation, anomaly detection, information leakage, as well as the deception of attackers, network jamming, or communications eavesdropping.

**Table II. Comparison of diverse FCS modeling techniques**

| Modeling Technique | Advantage | Disadvantage |
|---|---|---|
| Hierarchical Model (e.g. Stackelberg) | Can optimize simultaneously diverse system parameters | Exposures private data |
| Evolutionary Model | Suitable for systems in which players via trial and error learn that some strategies are better than others; this process is repeated until the evolution converges to a stable state among the players of a specific generation; this stable state represents the best response for each player. | High convergence time to reach a stable system state |
| Cluster-Based Model | Reward is shared among the elements of the same cluster | Not suitable for dynamic systems due to the high complexity to manage clusters |
| Differential Model | Suitable for dynamic systems, meaning the players' incentives to make their choices (i.e., select their actions) change during the time the game is played; the model optimization is done using a group of differential equations. | For games with more of two players, it could be difficult to evaluate the conditions under which such games have a Nash Equilibrium in the given class of player strategies, because each player lacks complete information about how the game was played (i.e. the game trajectory, the payoffs received) by the other players; consequently, the optimization of differential games with N players with incomplete information needs more research. |
| Potential Game Model | The same function runs in each node to optimize system configuration | A local optimum could be found instead of the expected global optimum |
| Mean Field Model | Can address large-scale and heterogeneous scenarios | Players are rational, indistinguishable, and influenced only by the average behavior of others |
| Stochastic Model | Dynamic repeated games that allow players to learn reinforced strategies towards system goals | It is very challenging to timely discover the equilibria of a stochastic game due to its random characteristics and large model dimensionality |
| Games-of-games Model | Cross-layered model design where in each | New utility functions are needed to represent in a formal and realistic ways the |

|  | system layer is played a game with features better adjusted to both the main functionality expected and the involved players | interdependency among the diverse games; other open issue is to discover the holistic equilibria of the games-of-games model, considering the bounded rationality of players, similar to what was proposed in [111] |
|---|---|---|
| Non-Cooperative Model | Highly suitable for dynamic distributed systems with incomplete information | Players cannot learn (with no game repetition nor cooperation incentive) from past actions |
| Auction Model | Low convergence time because the solution space is further reduced than in the case of GT | Every player should truthfully tell the system designer its intent; otherwise the model result is negatively affected |
| Gamification | Provides immediate feedback to aid players adjust their learned skills about a specific context | Replaces other learning activities such as simulations, creating uniform learned skills, reducing the diversity and eventually impairing the "Darwin evolution of skills" |

Performance Metrics for Fog Computing Systems

Table III summarizes a selection of performance metrics (or data labels in the context of machine learning) for FCSs, which deal with a heterogeneous resource set formed by computational, communications, storage, and energy assets. These FCS performance metrics can be classified into several types such as data, control, energy, resource allocation, processing, and business. All these metrics can be also analyzed and optimized by theoretical algorithms, among other possible ways, such as [112].

**Table III. Summary of selected performance metrics (or data labels) for FCSs**

| Metric (or Data label) | Main goal |
|---|---|
| Rate | Maximizes the access rate for a dataflow |
| Edge/fog caching | Proactive caching based on data/service popularity to minimize the access delay to data/service |
| Privacy | Level of data/service disclosure to unknown recipients |
| Trust | Degree of confidence in a data block/service by non-source recipients |
| Localization | Data/service delivery based on consumer localization |
| Fusion | Level of data aggregation and summarization |
| Reasoning | Level ok knowledge (labels) extracted from data features |
| Control delay | Selects the controller for an SDN-based node that minimizes the delay on the control loop |
| Energy | Maximizes the energy efficiency |
| Distributed resource scheduling | Allocates in a fair way scarce resources among competitors with distinct requirements |
| Computation offloading | Decision on deviating heavy processing tasks from constrained end-devices to more powerful edge servers |
| Parallel processing | Allows the division of an original high-complexity processing task into distinct subtasks, which are executed in parallel by edge servers |

| Processing delay | Minimizes the processing delay of a computational task |
| CPU/GPU capacity | How many operations CPU/GPU can process in a time interval |
| Processing cost | Minimizes the processing cost of a computational task |
| Virtual micropayment | Anonymous dynamic payments to incentivize the share of heterogeneous resources among players in distinct network settings |

*3.4. Novel Programming Approaches to Enhance the Management of Fog Computing Systems*

This subsection discusses relevant work associated with some recent approaches, such as SDN, NFV, or ML, which can enhance the management of an FCS. An FCS requires the deployment of sensors, actuators, and computing devices at the network edge. FCSs also need to be supervised and controlled de-centrally because of their complexity, heterogeneity, and geographical dispersion. To ensure network-wide resilience, it is fundamental to study the efficient orchestration [113] of a set of SDN applications that must cooperate among them to fulfil the global resilience requirements [114]. These have distinct goals including traffic classification, anomaly detection, or traffic shaping [114].

Software-Defined Networking / Network Function Virtualization for Fog Computing Systems

In parallel, the Quality of Experience (QoE) of end-user (or end-machine) services should be also supported end-to-end, among diverse network domains. To address these requirements, within each network domain, a Software-Defined Networking (SDN) [115] system with three levels can be deployed. The top level is formed by Network Function Virtualization (NFV) services and a northbound Application Programming Interface (API). The intermediate level is composed by an SDN controller or multiple SDN controllers and Southbound API. The bottom level incorporates networking devices and agents associated with end-user terminals or end-'things'.

The SDN controllers can support Quality of Service (QoS) only within a network domain. Although the SDN controller already has some abstraction from the hardware, that abstraction is limited, because the SDN controller typically contacts the OpenFlow switch-based devices and not the end-devices. Therefore, in these conditions it is very hard to support user-perceived QoE. Considering these limitations, the communication over the network domains should be made at the top level, by means of east-west APIs used by either SDN controllers or NFVs. In addition, the end-devices should report to the upper system layers (e.g. control and/or management) relevant statistical information about the services in operation within those devices. In this way, using the top-level, there is a higher abstraction level from the network infrastructure and the end-devices, and it enables powerful management interactions among the domains to assure a reliable end-to-end QoE. We argue that this kind of application-level programmable inter-domain environment is like the one previously proposed in [74]. Nevertheless, there is at least an important difference in the service needs between the previous work and future scenarios. In fact, the former is focused in objective QoS requirements [74][75] and the latter is more focused on subjective QoE requirements [53]. Thus, the service providers are now shifting their attention from intra-domain QoS fulfilment to end-to-end (and inter-domain) customer-perceived quality (i.e., QoE).

The literature offers several pieces of work that apply SDN for Moving Target Defense (MTD) network protection [116] in Fog Computing Systems. If a Fog Computing System is protected via MTD, then that system recurrently modifies its configuration to deter potential inspections by attackers into the ways the system is configured and operated. The authors of [117] propose to enhance SDN with NFV against penetration attacks. A penetration attack is perpetrated by a delivery method that transports a malign payload to the target system device. This malign payload can trigger on the target system device the execution of compromised code which can jeopardize the system normal operation. To circumvent the occurrence of a penetration attack on a specific system, that system should be comprehensively assessed to identify weaknesses that an attacker could exploit. This verification of potential system vulnerabilities is designated as penetration testing. In addition, the literature has a considerable number of SDN-based solutions to enhance diverse network features, such as network security [118][119][120], network communications [121][122][123], energy efficiency [124] and network lifetime [125].

The authors of [126] abstract the complexity of the physical world and present to a programmer an abstracted view of that physical world. Thus, the programmer can more easily create a system model, perform system debugging or explore the design space of various IoT applications. The abstracted view of the physical world is made by using the composition of several "accessors", which are design patterns that serve as proxies for any 'thing' or service that may be either local or remote. The accessors offer a similar functionality to that offered by a web proxy, when a client, instead of downloading a web page from the remote web server, alternatively downloads it directly from the web proxy. The work in [93] reviews and discusses adaptation features for deploying scalable and autonomous communication systems by means of SDN and NFV, both enhanced by ML.

Artificial Intelligence for Fog Computing Systems

FCSs are challenging to manage because of their high complexity. The complexity is due to both the internal operation of each system and the interdependence among systems. A possible way of managing these systems efficiently is to adopt some automatic control functions from natural systems that evolved to optimum operation modes with minimum energy consumption, ensuring the survivability of the species coevolving in those natural systems. Aligned with these important goals, artificial life (AL) is a research area that investigates natural systems related to chemistry and biology fields. According to the tool type used to perform the investigation in this area, there are three main types of AL: i) *soft* via simulation; ii) *hard* via actuators or sensors; iii) *wet* via biochemistry. We think AL techniques can be very powerful for evaluating and manage cyber-resilience in FCSs that show a highly dynamic behavior. A comprehensive coverage of AL techniques combined with self-organization system capabilities is in [127]. The evolutionary trend of AL is discussed in [128] and its relevant open issues are covered by [129].

Another area that can be easily associated with AL is artificial intelligence (AI), including deep learning [130][131], random neural networks [132] or deep learning with reinforcement learning [133], hierarchical learning models [101] or a very recent model designated as brain intelligence [134]. These contributions propose deep learning and other machine learning (ML) techniques to autonomously and optimally configure future wireless networking environments based on the information learned from network system behaviors. Those ML techniques optimize the model performance of high-complexity systems in a more efficient way than other legacy alternatives. This efficiency gain occurs because ML techniques discover the optimum results of a system model with smaller convergence time, higher accuracy, and are better adjusted to significant system variations than other legacy techniques that optimize the same system. Another limitation of a legacy technique is it does not support learning, in opposition to what ML methods guarantee [130]. The authors of [133] propose a secure and intelligent architecture for 5G mobile networks and beyond to enable flexible and secure resource sharing. Then, they suggest a Blockchain empowered content caching problem to maximize system utility and they develop a new caching scheme by utilizing deep reinforcement learning. A comprehensive survey on deep learning proposals for mobile and wireless networking is available in [131].

The authors of [135] surveyed the state-of-the-art applications of ML in wireless communication and they pointed out several unsolved issues. They divided the body of knowledge into resource management in the MAC layer, networking and mobility management in the network layer, and localization in the application layer. In addition, they identified several conditions for applying ML to wireless communication for aiding interested researchers decide whether to use ML and which kind of ML techniques to use. Further, the authors summarized traditional approaches together with their performance comparison with ML based approaches, based on which the motivations of previous work to adopt ML become more evident. For example, ML enables adaptive learning and intelligent decision making in mobile wireless networks, due to the ML capability to achieve the convergence the automatic processing of very large and complex input data sets and the systematic enhancement of self-adaptive algorithms [131][135]. All this is obtained by avoiding the manual preprocessing code obliged by rule-based and legacy techniques based on non-learning algorithms.

A tutorial on artificial neural networks-based ML for wireless networks is available in [136]. They discuss ML solutions to provide intelligent wireless networks and realize the full potential of 5G (and beyond) mobile networks.

The learning based on ML techniques in how to improve the system performance is normally obtained incrementally, during a considerable long-term period, and using methods that do not take in consideration the underlying engineering principles in arriving at their final decisions. Considering what the algorithms have learned (i.e. a set of weights to take the more suitable final decision) to improve the system performance, we can designate it as artificial intelligence that can automatically manage that system but, in a way, that it could be supervised by humans. In this area, the authors of [134] suggest brain intelligence (BI) as a new technique in AI to solve some optimization problems that cannot be solved by other, weaker AI algorithms. They discuss BI usage on scenarios such as autonomous vehicles, healthcare, and industrial automation.

The authors of [132] investigated the routing optimization for software defined networks even in severe use cases. They aim to optimize the QoS of data flows using a cognitive routing engine. Nevertheless, the satisfaction of QoS requirements can be jeopardized when system resilience / security is missing, or even when the system has other limitations such as limited resources of energy, which is very relevant for scenarios involving IoT devices. A new decentralized random-access algorithm is introduced in [55] to schedule the Plug-in Hybrid Electrical Vehicles (PHEV) charging to protect the distribution grid from bus congestion and voltage drop, and improve the grid efficiency. Another way to deploy learning is via either a hierarchical model using Stackelberg games [101] or deep-learning [137]. Further work is clearly needed in this field.

The current subsection highlights the great importance of investigating intelligent mechanisms to enhance the next generation of FCS [138] with new capabilities, e.g. self-awareness of resilience against threats. These new system capabilities can guarantee appropriate performance levels in dynamic scenarios, offer energy harvesting, diminish the consumption of energy, detect and recover from system errors, and protect against cyber-attacks. The self-aware management we have just discussed is like the autonomic (self-organized) management of networked systems, which is investigated in [139].

The next subsection discusses relevant factors for enhancing the resilience of FCSs.

*3.5. Key Aspects for the Resilience of Fog Computing Systems*

FCS resilience depends on several key aspects [140]. These are managing complexity, choosing the correct topology, adding redundant resources, designing for rapid recovery from failures in a distributed system, controlling failures and threats, providing adequate information buffering, deploying agents to enforce resilience, and analyzing system menaces.

<u>Managing Complexity</u>

As network complexity increases, then the network's resilience may be reduced, because the failure of a specific network component may cause the failure of other components in a completely unexpected way. The last unpredictable network behavior may be caused by some unplanned paths within the network that were not recognized by the network administrator or designer. Such unexpected behavior is particularly relevant in multi-genre or interdependent networks. These networks have also distinct roles such as data communication, computing, data storage, or extract knowledge from raw data. Therefore, unless a high level of complexity is needed to support resilience functions directly, the network complexity level should be controlled or even reduced. Aligned with this context, the authors of [141] discuss the existence of dependencies in complex systems and how the effect of those dependencies in systems operation should be characterized and analyzed. The authors of [142] present a methodology to assess the cyber resilience of a system controlling a specific geographical region. Their methodology can perform a system functional diagnosis identifying system parts that must be protected against cyber-threats, such as datacenters and communications networks. Otherwise, in the case when any of these system parts become exposed to a serious menace, the system performance can be seriously undermined.

Choosing the Correct Topology

The choice of the most adequate network topology used within a system can enhance the system's resilience [143]. In addition, there are two types of network topologies, according to the used node degree distribution in each network. The first type of network topology is exponential node degree distribution; the second type is scale-free network. Some examples of the former type are wireless networks and mesh networks and, of the latter type, the World Wide Web (WWW) and power grids. Comparing the two previous types of network topologies, one can conclude: i) on the one hand, scale-free graphs are much more robust to random node errors than graphs with an exponential degree distribution; ii) while, on the other hand, scale-free graphs are much more vulnerable to cyber-attacks targeted to some high-degree nodes.

Adding Redundant Resources

Providing crucial additional resources can improve the resilience of a system. As an example, within a power generation plant, when we increase the number of system nodes, the probability of a system failure can be reduced as well as a quicker service restoration after a system problem. In addition, the additional resources should have some distinct characteristics among them to create diversity within the system, avoiding the additional resources being affected exactly in the same way by a cyber-attack (e.g. a worm attack). So, combining new but slightly differentiated resources could create a more resilient system [144] but the system designer should be aware of the amount of system resources (e.g. energy, network nodes) being used to achieve the aimed level of system resilience. Considering again the energy consumption, the authors in [145] study a scenario of a vehicular wireless network with the goal of diminishing the energy consumption. To achieve this, they propose a solution that organizes the vehicles and other mobile nodes into clusters. In addition, to enhance further the energy efficiency they assume some collaboration among the nodes. Nevertheless, cooperating nodes may need to disclose sensitive information to others. This is an open issue that should be addressed in future work. In addition, [146] explores the controller placement problem in the context of software-defined Internet of Autonomous Vehicles to minimize energy consumption and support load balancing under latency limitations.

Designing for Rapid Recovery from Failures in a Distributed System

The correctness and performance of a fault-tolerant system depend on its underlying replication protocols. The authors of [147] propose a hybrid replication protocol that provides the high performance of memory-durable techniques while offering strong guarantees including disk-durable approaches. The key idea of their replication protocol is that the replication mode should depend upon the state the distributed system is in at a given time. In the common case, with many (or all) nodes up and running, the last solution runs in memory-durable mode, thus achieving excellent throughput and low latency; when nodes crash or become partitioned, the same solution transitions to disk-durable operation, preferably flash-based solid-state drive (SSD) disks, thus ensuring safety at a lower performance level.

Controlling Failures and Threats

The designer of a system should protect the system against cascaded failures. Such failures occur when a node failure triggers a neighbor node failure and so on. To avoid these sequential (cascading) failures, the dependencies among nodes should be planned to minimize the chance of a failure easily propagating via neighboring nodes [143]. In addition, the effect of human actions in the way a failure could propagate within a system should be also studied [148]. The work in [149] evaluates some state of the art anomaly detection mechanisms to assess their monitoring and detection features of a challenge or threat to the normal operation in multi-tenant cloud infrastructures. They have concluded that elasticity and live migration of cloud services impair the detection of the traffic normal behavior, and consequently its correct isolation from the traffic associated to any anomalous incidents that are likely to be initiated.

Providing Adequate Information Buffering

To offer robust and timely data access, despite scarce network resources, some resilient solutions based on network buffering are very attractive. As an example, network buffering has been used to restore link connectivity and network performance following topological changes in mobile ad hoc networks [150], as well as to diminish the data access delay in disruptive-tolerant networks [151]. A major origin of packet loss is related to the classical problem of congestion and limitations in network resources (e.g., link bandwidth, router buffers). In order to prevent network congestion, the network administrator could decide on the proactive oversubscription of network resources. However, this strategy results in a significant amount of network resources not being used when the network load is low, increasing the cost of the network operation, which in some cases will not be acceptable. Thus, there is a tradeoff between management actions to avoid loss of messages and the associated costs. A possible way ahead is to use machine learning techniques [136] to detect the possibility of network resource contention as early as possible. If the resource contention can be identified early enough, then packet forwarding through the network can be dynamically adjusted so that congestion and thus message losses are avoided even at high network loads.

Deploying Agents to Enforce Resilience

To allow a system to absorb a specific failure or attack, recover from that issue, and adapt the system for mitigating that problem in the case it occurs again, it is necessary to deploy active agents, either human or artificial. When the active agents enforcing the system's resilience are humans, they should be trained, prepared, and motivated to perform the functions of absorbing, recovering, and adapting from an eventual failure, as efficiently as possible [148]. Alternatively, artificial intelligence techniques [152] can be used to deploy artificial agents that, on one hand, carry actions to enforce the problem mitigation and the system recovery and, on the other hand, the same artificial agents maintain a required level of concealment, exercising a self-defense strategy against the adversaries, e.g. malware that aims to discover and destroy the artificial agents of a system.

Increasing the functional redundancy within the agents that manage a network can significantly enhance the resilience of network functions in case of a network perturbation, e.g. loss of some agents. Using functional redundancy, the system roles made by the lost agents are reallocated to others. Another advantage of having several agents performing the same function is increasing the system's scalability against high levels of system demand. The authors of [153] describe a distributed decision algorithm supported by diverse SDN controllers to enhance the recovery mechanism from a problem. The authors of [154] review comprehensively the major principles and challenges for the design of smart Fog Computing Systems that can recover at run-time from unexpected faults or threats.

Analyzing System Menaces

The system attacker specifically tries to defeat the absorption and recovery efforts of the resilience strategies in order to perpetrate the worst impact possible on the system's normal operation. In this way, the system designer should protect the processes of absorption and recovery after a threat so that these processes are less penalized by malicious actions. The work in [155][156][144] propose GT to identify effective strategies against cyber-threats.

Any resilience-enhancing measure can cause unanticipated effects, leading to an overall reduction in system resilience. Therefore, each resilience-enhancing measure should be analyzed to check if it could have a potential negative effect on the system operation. Thus, comparative analytical studies should be made with and without the measure being investigated. Ref. [157] offers a numerical resilience definition that allows system designers to assess in a more formal way how much the system resilience could change after a set of system alterations are performed. It can also compare the resilience between either distinct systems or various design options of the same system.

In addition to the relevant factors to enhance the resilience of Fog Computing Systems, which were discussed in the current subsection, the reader can consult [13] to obtain an alternative set of attributes that a system should satisfy in order to become more resilient against diverse kinds of system faults namely errors, failures or attacks. Further, [26] classifies typical faults present in a SDN-based system.

The following Section discusses a proposed design to enhance the operation of legacy Cypher-Physical Systems (CPSs) by deploying the technologies and other aspects previously discussed in the current Section towards a new generation of more resilient CPSs, which we generically designate in this paper as Fog Computing Systems.

**4. Hierarchical Design of a Resilient Fog Computing and IoT System**

This Section discusses design aspects that are important to consider in a resilient Fog Computing and IoT System. Table IV presents a four-layered hierarchical architecture that can detect, absorb and recover, and adapt to threats against Fog Computing and IoT Systems [8]. The bottom layer of Table IV measures, collects (typically inside a local IoT domain), and stores data obtained via either physical or virtualized devices. The detection and absorbing of threats against a Fog Computing System (FCS) is made via interface chip programming.

The second layer of the architecture visualized in Table IV aims to distribute a common pool of available resources, such as, communications, storage, processing, or services. This layer is typically responsible for a single intra-domain using devices that exchange data using a single communications protocol. The exchange of data is controlled by flow rules stored within local devices and queues to give differentiated QoS to flows. Alternatively, the QoS support can be deployed based on flow type (e.g. video streaming, interactive, best effort), enhancing the system scalability, because the same flow type can aggregate many individual traffic flows.

The third layer of Table IV aims to control the data plane topology and how the traffic should be transferred through the available topological links. This layer is also essential to support the feedback loop between the cyber and physical worlds. The current layer uses SDN-based solutions to control a heterogeneous intra-domain communications infrastructure, formed by diverse area networks with distinct networking ranges, such as home, large building, or corporative enterprise. The normal operation of each area network can be protected via an SDN-based framework as suggested in [158]. This framework creates a secure perimeter around a network domain, protecting it in a completely distributed way from the negative impact induced by several types of cyber-attacks. A possible enhancement on this framework can be achieved by modifying it to also detect and mitigate any functional failure or even functional degradation on the network topology.

The fourth and topmost layer of Table IV is the management layer. It supports service discovery, service composition, service management, and service interfaces. First, the service discovery aims to verify if a required service is available within a specific system. Second, the service composition aggregates diverse services, coordinating among them, as if these services operate has only a single service. The service integration is useful in a scenario where to satisfy a specific system request then several system services need to be processed for that request in a pre-defined order. Third, the service management manages and determines the trust mechanisms to satisfy in a resilient way the service requests involving diverse internal services. Fourth and finally, the service interfaces are used to support interactions among all the provided services within the management layer. The last layer aggregates several approaches, such as NFV and SDN. The current layer also deploys a heterogeneous inter-domain communications infrastructure, formed by diverse area networks with distinct networking ranges, such as street, digital city, or wide area.

The topmost layer (Cyber, Social) of our proposed architecture supports the functional requirements of either human users or agents. The application layer supports applications used within FCSs with embedded sensors and actuators including smart grid, smart transportation, smart cities, smart homes, smart farming, smart health care, smart logistics, or smart industries. Most of these applications are completely distributed, among diverse networking domains, which brings new security challenges. To address these security challenges, we envision several ways, using middleware [159][160] along with other more recent options, such as Blockchain smart contracts [161][162]. These solutions should offer a secure management functionality to ensure a resilient and well-coordinated FCS operation. Using smart contracts, it is possible to deploy trustful roaming (peer to peer) services among nodes belonging to distinct network domains, avoiding the use of external communication entities to enforce security, such as key distribution centers. Another advantage of using distributed smart contracts is higher resilience against network failures. The authors of [163] proposes a smart contract for collaborative edge learning, ensuring authentic and correct message exchange during the training process, and detouring the behavior of malign entities. The modelling of Blockchain-based proposals is comprehensively discussed in [164].

The requirements of emerging applications can be specified in YANG service modules [165], including a description of how each application is expected to be experienced by the customer. To give an example, a service model can be associated with video traffic. Then, YANG service modules are processed to identify the (network, processing, storage, etc.) intents [166][167]. For example, a network intent could be "optimize my network for enhancing video QoE". Then, the intent is translated to a management policy. This specifies a logical (Boolean) condition to fire an action. The action is defined independently of the networking device that will be deploying that action. This action could be "send the received message to the more suitable next-hop interface to the final destination". The next and final step is to convert the management policy to specific device rules. Performing this final conversion, the management policy action "send the received message to the more suitable next-hop interface to the final destination" is converted to some specific actions on the data plane device. In this case, supposing a device such as an SDN-based switch, the field action of that device rule could be, "Output: port 2", where "port 2" is the switch port number two.

The initial management intents and/or policies can be adjusted (or indeed augmented) by Machine Learning (ML) [168] algorithms. These algorithms learn from the system operation and build a sort of Artificial Intelligence (AI). This self-awareness capability is possible after a set of consecutive successful adjustments on the system configuration. A system configuration adjustment could occur after the system being in operation, during a specific time interval (i.e., epoch), using the previous system configuration. In this way, the top-level layer can collect the system statistics, during the previous time interval, and deliver the statistics (e.g., using input attribute weights) to ML/AI algorithm. Then, this algorithm for indirectly improving the video QoE (e.g., the video QoE is represented via a function output) can decide to notify the Intent Engine (e.g., using output attribute weights). Thus, the Intent Engine changes from the current management policy to another one. This new management policy is then converted by the SDN Controller to a set of novel rules, which are pushed via a Southbound API to the data plane devices being used by the optimum video data path through the network infrastructure. The functionality which we have just described can be summarized such as an observation (i.e., collect of statistics) – action (i.e., set of novel rules for the data plane devices) loop. This new type of observation-action control loop, made possible via the SDN paradigm, can be seen as a promising basis for incorporating in the management of future networks some powerful self-driving capabilities [93]. This new management of future networks will be completely opposed to the closed-form models of individual protocols [169], such as the control of network congestion by active queue management [170]. In fact, the management of a self-driving network, in opposition to the legacy management of network resources, implies the next new aspects [169]: i) the network measurement is task-driven and tightly integrated with the control of the network; ii) the network control relies on large-scale analysis over the acquired data from the global operation of the networked system; and iii) the major outcomes from the last analysis form an important system learning asset to orchestrate and efficiently control the network available resources.

**Table IV. Hierarchical Design of a Resilient Fog Computing and IoT System**

| Layer | Plane | Domain | FCS Activity [8] | Goals | Tools |
|---|---|---|---|---|---|
| 4 | Intelligent management (Cyber, Social) | Inter/Intra | Adapt | Reasoning, orchestration, full abstraction, Yang service models, intents, management policies | NFV, SDN, Intent Engine, ML/AI |

| 3 | Control | Intra | Adapt, recover | Partial abstraction, topology, traffic, control actions, system status | Software-Defined Controller with link layer discovery, forwarding, and feedback loop |
| 2 | Switching | Edge | Detect, absorb, recover | Decision about next link decision, traffic mirroring, discard packet | OpenFlow rules in local device tables, queues |
| 1 | Sensors, Actuators (Physical, Virtualized) | IoT | Detect, absorb | Accept or discard received message | Interface Chip programming |

The next Section debates the lessons learned along the current paper and open research issues.

## 5. Lessons Learned and Open Issues

This Section summarizes and highlights the main findings from the analysis and reflection in our paper. We start by debating interesting ways to enhance the management of resilient FCSs. Then, in subsection 5.2 we discuss the deployment of network slicing proposals to empower resilient FCSs. Subsection 5.3 deals with the promising area of data fusion for resilient FCSs. In addition, subsection 5.4 discusses literature work related to proactive and preventive maintenance of systems that we envision could be also applied with great success to resilient FCSs. Then subsection 5.5 debates the relevant future need to efficiently disseminate data in resilient FCSs. Finally, subsection 5.6 investigates the integration of Blockchain and Machine Learning for resilient FCSs.

*5.1. Interesting Propects on the Enhanced Management of Resilient Fog Computing Systems*

We consider now interesting aspects for further improvement on the management of FCSs in hazard scenarios. We envision the design and programming of intelligent solutions that provide and optimize the autonomic management of interconnected heterogeneous FCSs, orchestrating the available system resources in an abstracted, elastic, flexible, and stable ways. This new context can support the model of detect-absorb-recover-adapt discussed in Section 4 [8], which can be used towards a resilient systems-of-systems [9]. This can be successfully applied to various FCS scenarios with networking infrastructures (see Table V) for fulfilling the goals of energy efficiency [124], quality provision [171], computation offloading [105], mobility support [172], data fusion [173], data offloading [174], threat management [175], and online optimization for distributed fog networks [176]. We detail below these relevant goals to enhance the robustness of FCSs and consequently enhance their performance in adverse situations. Considering that many devices in FCSs are battery-operated, then the aspect of increasing the energy efficiency in those systems becomes very relevant. When the available energy in each battery is used in a more efficient way the system operational lifetime is increased. In addition, the system sustainability is improved.

**Table V. Applying the model of detect/absorb and recover/adapt** [8] **to various FCS scenarios**

| Scenario | Goal | Detect (event-based) | Absorb and Recover (Control) | Adapt (Learn) |
| Mesh network formed by IoT battery-operated devices | Routing with energy efficiency | Detect network devices with a low-level of battery via associated events | Do not select routing paths using network devices with low energy in their batteries | Balance the traffic load through diverse paths to guarantee a fair depletion of battery among all the network devices |
| Quality provided by network slicing | To guarantee the delay of a specific flow type is constrained to a maximum | Event originated by a high packet delay | Select alternative path or discard some packets | Based on historical data analysis performs a proactive load routing |
| Computation offloading from end-user devices to edge servers | Efficient usage of end-user device with limitations on computational resources | Events originated by high CPU utilization at end-user devices | End-user device with maximum CPU utilization during a specified time alleviates its burden by moving the execution of some tasks to servers at the network edge | Based on historical data analysis, performs proactive actions of computation offloading |
| Mobility support | Guarantee a pervasive access to data and services | Events originated by end-user mobility, active message flows, and available resources | Offer a seamless flow handover among distinct mobile access networks | Based on previous learned mobility behaviors, disseminates in advance through the edge of the network infrastructure several replicas of message flows and/or services |
| Data fusion | To shrink the high data volume from IoT devices to a much lower data volume, and reduce resources | Events from heterogeneous sensors transport raw data | Edge network devices aggregate and synthetize useful information from the received raw data | Based on historical data analysis, optimizes the methods to merge and extract useful knowledge from the received IoT data |
| Data offloading | Edge data caching based on spatio-temporal popularity | Events originated by high end to end Round Trip Time between a request and its reply | Stores data replicas at diverse edge network nodes | Based on historical data analysis, performs a proactive data replication at selected network nodes |
| Threat management | Mitigation of system threats | Events originated by threat detection; events are classified as "malign traffic" or "packet is lost" | If threat is "malign traffic" then "discard malign packets" elseif threat is "packet is lost" then "select an alternative and more robust path" | Based on historical data, the FCS can autonomously adjust the previous action against a future occurrence of the same (or similar) incident, and a trust indicator; the value of this trust indicator shows how well that action could mitigate the associated incident. |
| Computation offloading | Jointly optimize the formation of | Fog nodes events originated by | Offloading computational tasks to selected | To minimize the maximum latency when computing a new task that arrives at a fog |

| among fog and cloud servers | fog networks and the distribution of computational tasks in a hybrid fog-cloud system with dynamic fog nodes | their arrival, their mobility pattern, their localization, their local capabilities, their support of micropayments | neighboring fog nodes and the cloud | node under uncertain behavior of other fog nodes |

*5.2. Network Slicing for Resilient Fog Computing Systems*

IoT devices produce a lot of data demanding a large amount of network resources, and the best-effort allocation of network resources may not be efficient. To overcome this, it will be interesting to investigate novel ways to manage system resources, such as network slicing [171]. Network slicing is a technique that allocates to each data flow a dedicated set of network resources, according to the specific QoS/QoE requirements of that flow. To scale out the system management solution instead of allocating network resources to individual data flows, the resources should be allocated to data flow types (e.g. video, gaming, best-effort). This is like the classic DiffServ QoS strategy that (per domain) classifies, marks, polices, and shapes the incoming traffic class. In contrast to DiffServ, network slicing can be used to meet the demands of vertical applications in 5G networks, across multiple domains, end-to-end, as discussed in [177]. To realize the network slicing goals, we need software that automates the creation, supervision, and deployment of resources and services in the underlying virtualized context. This software is normally an orchestrator of both services and system resources. Ref. [177] discusses open source orchestrators for network slicing, such as OpenMANO, OpenBaton, OPNFV, M-CORD, or ONAP, among others.

End-user devices have serious limitations on computational resources. In this way, some important system performance gains can be obtained as some heavy-processing tasks are moved from the resource-constrained end-user devices to more powerful servers at the network edge [105]. This distributed management of tasks offers the simultaneous advantages of finalizing quickly those tasks and increasing the battery autonomy of battery-operated end-user devices. In the opposite direction of what we have just discussed, [178] debates future directions in networked control systems, where the execution of delay-sensitive tasks, such as the control of robotized arms, can be automatically offloaded from remote systems to fog controller nodes, giving significant gains in the control accuracy of field devices.

The authors of [172] propose a solution based on network slicing to support seamless flow handover among mobile access networks. Nevertheless, they acknowledge that further research work is necessary in providing end-to-end network slicing to address the following open issues: network reconstruction over heterogeneous technologies, the high complexity and difficulty of slicing management, and the lack of cooperation with other 5G technologies such as C-RAN, SDN, and NFV. Further related research directions are in [179]. As these open issues will be successfully endorsed, some interesting and relevant functional outcomes can be obtained namely seamless mobility, high transmission rate, ultra-low guaranteed latency and jitter. These aspects are essential to support robust loop controls through the future data communications infrastructures, which should support both a reliable, efficient, and timely exchange of data among machines, agents (algorithms) or end-user devices.

A very recent technology designated as Segment Routing for IPv6 (SRv6) has been investigated in both the standardization and research communities to build novel distributed processing models on top of the network layer, and for various networking deployment scenarios [180]. The authors comprehensively survey and discuss segment routing, including SRv6. They also point out possible future research directions, and domains in which segment routing would bring benefits, namely: i) service function chaining support; ii) srv6 end-host implementation aspects; iii) cloud orchestration; iv) mobile 5G; and v) IoT.

*5.3. Data Fusion for Resilient Fog Computing Systems*

Due to the high level of heterogeneity and complexity of diverse Fog Computing systems as well as the need for a high-level of interoperability, management solutions with "fusion-enabled" capabilities [86][181][173] have been recently proposed. These use distinct system architectures such as the Internet and Information Centric Networking (ICN) for efficiently sharing common resources via interoperability entities [86]. Alternatively, the same fusion-enabled capability can be applied to merge and process heterogeneous data from multiple sources, leading to better estimation accuracy of the status of managed physical systems [181][173]. The data fusion methods can be classified based on the data space of each use case, namely Fog Computing space fusion, Cyber-Social space fusion, or Cyber-Physical-Social space fusion [181]. Considering the most complex case, the authors of [181] discuss a Cyber-Physical-Social System that aims to predict multi-users' mobility pattern by solving a cubical User-Spatio-Temporal probability arising from heterogeneous sensor data. To solve the stationary probability map, they refer to a tensor-based iterative algorithm to merge and process sensor data from multiple sources, namely time, space, and social network. The algorithm seems promising for the prediction accuracy and associated convergence time. In addition, other important aims are to perform data fusion securely as well as preserving data privacy [173].

*5.4. Proactive and Preventive Maintenance of Resilient Fog Computing Systems*

Future research in high-level preparedness against threats to FCSs should certainly include proactive and preventive maintenance using either discrete-time [182] or continuous-time control [175] of IoT systems, incident protection [111], short mitigation times, and fast recovery. It is also very important to pay more attention to the scenario of multiple threats simultaneously affecting the normal operation of an FCS. The diverse aspects to be enhanced should be studied not only at system runtime but initially at its design [12], including the protection of the physical infrastructure against weaknesses recently reported [6]. Moreover, the FCS design enhancement should not only consider the protection against threats but other relevant aspects such as the quality of service offered to system clients and the energy consumed by system [183].

*5.5. Efficent Data Dissemination in Resilient Fog Computing Systems*

The data transfer is exponentially increasing through the communications infrastructure of FCSs. Nevertheless, the fast transfer of data can be adversely affected by high mobility and congestion, essentially in mobile networks. To diminish the data transfer time, the authors of [174] propose a solution that estimates the content popularity. Then, that content popularity is used to decide about which content is offloaded from remote clouds and cached at Base Stations to achieve higher user satisfaction and backhaul offloading. As future work, the content popularity can be evaluated combining several data-related dimensions such as spatial, temporal, mobility behavior and resource availability, including energy [184].

Future networks should provide on-time delivery of flow packets, respecting latency constraints, i.e. delay always in the range, [delay_min, delay_max], for each application [39]. This new behavior should be a distinct improvement on current network services where service quality may vary according to the network load. The TCP transport protocol should be also revisited, specifically its algorithm for controlling end-to-end congestion together with the right buffer size for routing devices [185].

Emerging applications will require the simultaneous satisfaction of several requirements, in very precise and scalable ways, especially those of throughput and latency. However, resilience should also be fundamentally guaranteed for applications that demand it, specifically for remote surgery or autonomous vehicle driving applications. Further, urgent, investigation is required in these areas.

*5.6. Blockchain and Machine Learning Integrated Solutions for Resilient Fog Computing Systems*

As was discussed in subsection 3.1, PKI is a hierarchical centralized model that validates the authenticity of digital certificates. This centralized management of networked resources presents several limitations including the low level of offered resilience against system threats. Alternatively, management solutions that rely on more decentralized and distributed operating modes seem preferable than their centralized counterparts for managing resilient computational resources of both remote and fog networked infrastructures. To guarantee all these conditions in a scalable way, the deployment of solutions involving Blockchain seem promising [186]. Nevertheless, popular Blockchain solutions such as Bitcoin and Ethereum, which are data-based, are often limited by scalability challenges and latency in transaction processing, due to the scarceness of computing resources at the network edge to guarantee a universal consensus among miners towards the final decision about how to update the centralized chain. To mitigate the limitations of data-based virtual currencies, there are other alternatives based on agent-centric architectures such as Holochain [187]. It enables any device to have its own chain-based ledger system. This allows every device on a system to function independently, and requires data synchronization only when necessary, e.g. to support data redundancy which is useful to give resilience against an agent failure. The sharing of data among agents is made via a distributed hash table. In this way, Holochain can support more scalable distributed applications with data integrity that can run in mobile battery-operated terminals, which are normally constrained in terms of computational capabilities and available energy. In addition, other possible solutions in the literature for how the system should be managed in a resilient manner are auctions [188] or innovative business models [189].

Other work [190][191] proposes not only Blockchain but its combination with machine learning to manage edge computing scenarios with IoT devices more scalably and efficiently. The convergence of edge computing and deep learning is comprehensively investigated in [192]. The authors discuss the scenario of deploying deep-learning algorithms at the network edge to enforce intelligent ways to manage the resources available at the network periphery among the diverse system nodes. In addition, Blockchain (or Holochain) can be used to support trust among the edge devices as well as giving enough incentives to their cooperation towards common system objectives. Further research is needed to find solutions for the goals of [192]: i) executing complex deep learning computations; ii) supporting the live offloading of microservices to reduce service latency, energy consumption, or service unavailability; and iii) orchestrating assets among the cloud and distributed edge servers to achieve better network operational performance.

## 6. Conclusion

This paper has explored the provision of future Fog Computing Systems (FCSs) from the perspective of the literature on resilience properties for modern Cyber-Physical Systems supported by IoT and edge computing elastic resources. Our discussions covered key use cases, notably those of power grids, smart buildings, mobile networks, healthcare, and Industrial IoT. We reviewed work that aggregates GT, SDN/NFV, and multi-access edge computing for enhancing system resilience and maintaining normal system operation under distinct and serious threats. We also discussed system design and how SDN-based theoretical model algorithms can be used to optimize system operation with respect to well-identified goals such as energy efficiency, computation and data offloading, management of flow quality, and IoT data fidelity. Along the way, relevant topics for further research in this increasingly important research area have been identified.

**Funding:** The work of Jose Moura is funded by FCT/MCTES through national funds and when applicable co-funded EU funds under the project UIDB/EEA/50008/2020.

**Acknowledgments:** Jose Moura acknowledges the support given by Instituto de Telecomunicações, Lisbon, Portugal. Jose Moura is grateful to Beatriz Moura for her initial work on the paper figures. David Hutchison is grateful to many colleagues in the EU COST Action RECODIS (CA15127) for discussions on the resilience of communication systems. We thank the reviewers very much for their constructive and positive comments, which have helped improve the quality of this paper.

30 of 38